\documentclass[twocolumn]{webofc}

\usepackage[varg]{txfonts}   
\usepackage{hyperref}
\usepackage{url}
\usepackage{float} 

\hypersetup{colorlinks=true,citecolor=blue,urlcolor=blue,linkcolor=blue}
\begin{document}
\title{Results on \textsuperscript{235}U(n\textsubscript{th},f) isotopic fission yields using prompt and delayed gamma rays at the FIPPS spectrometer of the ILL }
%
%


\author{\firstname{Thomas}
\lastname{Materna}\inst{1}\fnsep\thanks{\email{thomas.materna@cea.fr}} 
\and \firstname{Pierre} \lastname{Herran}\inst{1}
\and \firstname{Mattéo} \lastname{Ballu}\inst{1}
\and \firstname{Quentin} \lastname{Poirier}\inst{1}
\and \firstname{Alain} \lastname{Letourneau}\inst{1}
\and \firstname{Diane} \lastname{Dore}\inst{1}
\and \firstname{Loïc} \lastname{Thulliez}\inst{1}
\and \firstname{Caterina} \lastname{Michelagnoli}\inst{2}
\and \firstname{Felix} \lastname{Kandzia}\inst{2}
\and \firstname{Yung Hee} \lastname{Kim}\inst{2}
\and \firstname{Ulli} \lastname{Köster}\inst{2}
\and \firstname{Olivier} \lastname{Litaize}\inst{3}
\and \firstname{Olivier} \lastname{Sérot}\inst{3}
\and \firstname{Abdelaziz} \lastname{Chebboubi}\inst{3}
}

\institute{ IRFU, CEA, Université Paris-Saclay, 91191 Gif-sur-Yvette, France 
\and
           Institut Laue-Langevin, CS 20156, 38042 Grenoble Cedex 9, France
\and
           CEA, DES, IRESNE, DER, Cadarache, F-13108 Saint-Paul-Lez-Durance, France
           }

\abstract{We present preliminary results from a measurement campaign conducted with the FIPPS $\gamma$-ray spectrometer at the ILL using an active target made of \textsuperscript{235}U dissolved in a liquid scintillator. Prompt $\gamma$ rays were used to extract absolute, independent isotopic fission yields. We obtained yields for a selected set of well-produced even-even nuclei from the \textsuperscript{235}U(n\textsubscript{th},f) reaction and compared them with JEFF-3.3 evaluated data. This included the doubly magic nucleus \textsuperscript{132}Sn, for which we observed a pronounced deficit (about a factor 7) with respect to JEFF-3.3. Using the FIFRELIN fission fragment de-excitation  code, we interpreted this discrepancy as evidence that \textsuperscript{132}Sn is mainly produced in its ground state at scission or after neutron evaporation. Delayed $\gamma$ rays recorded in the same experiment were also analysed. No deviation was found 
in the beta decay of $^{132}$Sn compared to the nuclear database. However, an excess of about a 40\% was found 
in the $\beta$ decay of \textsuperscript{90}Kr, which may originate from an overestimated ground-state feeding of $^{90}$Rb in the evaluated data. 
}
\maketitle
\section{Introduction}
\label{intro}
Independent fission yields are essential nuclear data for applications and for testing nuclear fission models. However, obtaining such data with good mass and charge resolution remains challenging. In thermal neutron-induced fission, the best results have been achieved with the LOHENGRIN fission-fragment mass separator at the ILL, but nuclear charge identification is only possible for Z less than about 40 \cite{langNuclearChargeMass1980}. Reverse kinematics experiments performed with, e.g., SOFIA at GSI \cite{pellereauAccurateIsotopicFission2017} provide remarkable performances, but at the expense of poor definition of the excitation energy of the fissioning system.
An alternative, indirect approach for obtaining independent yields is to exploit the prompt $\gamma$ rays from fission fragments. This technique is promising, as fission fragments are identified unambiguously by $\gamma$ rays in their de-excitation cascade. It has already been applied to ion-induced fission reactions \cite{bogachevFissionFragmentProperties2007} and to fast-neutron-induced fission of \textsuperscript{238}U \cite{wilsonAnomaliesChargeYields2017}. It was also used for the well studied \textsuperscript{235}U(n\textsubscript{th},f) reaction \cite{mukhopadhyayPromptSpectroscopicStudies2012,deyMeasurementRelativeIsotopic2021}, but a comparison of the method with evaluated data was not fully carried out. 

In the first part of this article we present our attempt to determine precise absolute independent fission yields for a small set of even-even fission fragments for the thermal-neutron-induced fission of \textsuperscript{235}U. The measurement was performed with the FIPPS $\gamma$-ray spectrometer of the ILL \cite{michelagnoliFIPPSFIssionProduct2018}. The main difference with previous works  \cite{mukhopadhyayPromptSpectroscopicStudies2012,deyMeasurementRelativeIsotopic2021} is the use of an active fissile target \cite{kandziaDevelopmentLiquidScintillator2020} which allows us to separate fission events from beta decays and to normalize per fission. The aim of this analysis was to a large extent to verify whether the anomaly in the \textsuperscript{132}Sn yield reported in \cite{wilsonAnomaliesChargeYields2017} and later refuted in \cite{ramosFirstDirectMeasurement2019}, also appears in our data and, if so, to investigate its origin.

In the second part of the article we present initial results from our analysis of delayed $\gamma$ rays recorded in the same experiment. The context of this work is related to the reactor antineutrino anomaly \cite{mentionReactorAntineutrinoAnomaly2011}, which refers to the fact that reactor experiments observe a 5\% deficit in the antineutrino flux compared to prediction based on reference $\beta$ spectra \cite{muellerImprovedPredictionsReactor2011,huberDeterminationAntineutrinoSpectra2011}. Although the hypothesis of a sterile neutrino was rejected, the anomaly itself was confirmed \cite{thestereocollaborationSTEREONeutrinoSpectrum2023}. No compelling explanation has been established, apart from a possible bias in the $\beta$-spectrum measurements performed at ILL.    
A prediction of the antineutrino and $\beta$ spectra may be derived directly from nuclear data, but this requires improving decay data for fission fragments.  Some of these suffer from the Pandemonium effect and as consequently from an overestimation of ground state feeding. We therefore evaluate the possibility of extracting precise delayed $\gamma$-ray information from our data for a small set of nuclei, in order to test their consistency with evaluated decay data and cumulative fission yields. It may indeed be a faster way to identify large discrepancies in nuclear databases.
\section{Experimental campaign at FIPPS}
\label{fipps}
\begin{figure}[h]
    \centering
    \includegraphics[width=\linewidth,clip]{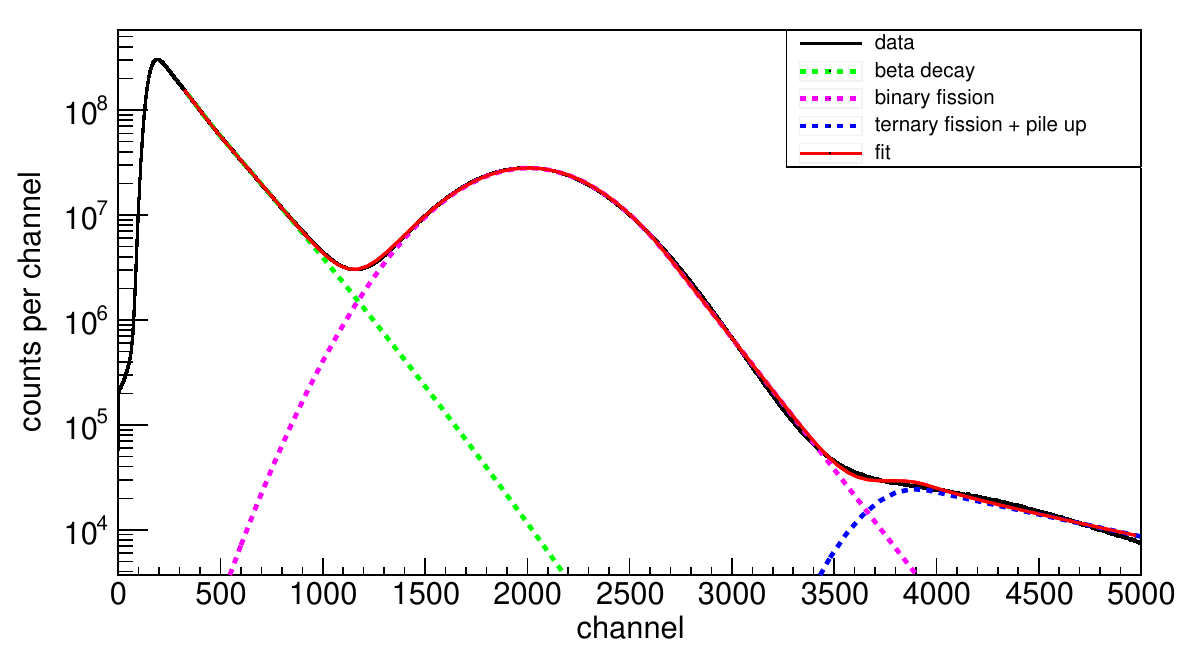}
    \caption{Spectrum of the photomultiplier tube detecting the light from the active target. Data are fitted with a simple model with 3 components: mainly an exponential function for the $\beta$ decay part, a double-shouldered gaussian function \cite{dasSimpleAlternativeCrystal2016} for binary fission and a double-shouldered gaussian function for ternary fission + fission pile up. }
    \label{fig-htag}
\end{figure}

The FIPPS $\gamma$-ray spectrometer, composed of 16 HPGe clover detectors, was installed at the end of a thermal-neutron guide of the ILL, where the 1-cm-diameter collimated beam flux was about 10\textsuperscript{8} neutrons.cm\textsuperscript{-2}.s\textsuperscript{-1} \cite{michelagnoliFIPPSFIssionProduct2018}. 
The active target consisted of \textsuperscript{235}U dissolved in an organic scintillator. Its working principle and performances are described in detail in \cite{kandziaDevelopmentLiquidScintillator2020}. When a fission occurs in the target, the two fission fragments are fully stopped in the liquid and their kinetic energy is converted into visible light. This light is then guided by a set of mirrors toward a photomultiplier tube (PMT). When a fragment undergoes $\beta$ decay in the target, the $\beta$ electron also deposits parts of its energy in the liquid. Despite the strong quenching of heavy ions compared to electrons, the amount of light generated by the energy deposition of the two fragments is larger than that produced by a $\beta$ electron (Fig.~\ref{fig-htag}). This allows us to discriminate the two processes and to sort any $\gamma$-ray detected by the spectrometer as either a prompt $\gamma$ ray (fission event) or a delayed $\gamma$ ray ($\beta$-decay event). 

In our analysis, we considered all the $\gamma$ rays in coincidence with a PMT  signal higher than a fixed threshold (here channel 1100) and within a 430 ns time window, as prompt $\gamma$ rays coming from fission. $\gamma$ rays outside this coincidence window, as well as in coincidence but with the PMT signal below the threshold, were sorted as delayed $\gamma$ rays. Fitting the PMT spectrum provides an accurate estimation of the total number of fission events:  $2.45(1)\times 10^{10}$, for the entire experimental campaign, or more exactly for the 9095 five-minute runs we selected for the analysis. The active target container is made of sapphire (Al\textsubscript{2}O\textsubscript{3}). The large number of thermal-neutron captures in aluminum and their subsequent decays explain the high proportion of $\beta$-decay events in the PMT spectrum.    
\section{First results on independent fission yields}
\label{prompt}
Since we have access to prompt $\gamma$ rays, the independent fission yield of a given fragment can, in principle, be determined by summing the intensity of all the transitions feeding its ground state and then by dividing that sum by the number of fissions.  This method (M1) is expected to work well for even-even nuclei. For those, only few transitions dominate and the analysis can be limited to fitting one or two well-identified peaks in the prompt $\gamma$-ray spectrum.  In practice, however, the prompt $\gamma$-ray spectrum is often too complex to reliably extract the intensity of a peak, due to peak overlaps. Indeed, it contains the prompt $\gamma$-rays emitted by all the fragments produced in the fission reaction. 

A solution may be to use the prompt $\gamma$-$\gamma$ coincidence matrix instead and to sum over all the possible cascades of two successive transitions ending in the ground state. The method (M2) we adopted consider only the most probable (and/or cleanest) such cascade in the fragment. The ratio of the two transition intensities, and the relative contribution of any other transitions to the ground state, are obtained by requiring a coincidence with two transitions in one of the (usually most probable) fission partner of the fragment. Mathematically, M2 is equivalent to M1 under the assumption that the fragment de-excitation cascade does not depend on the specific choice of its fission partner. Obviously, both methods rely on the prompt histograms ($\gamma$ spectrum, $\gamma$-$\gamma$ matrix, $\gamma$-$\gamma$-$\gamma$ cube) being free of contamination from delayed $\gamma$ rays, since the same transitions may be produced both in fission and in $\beta$ decay.
The M2 method was applied to nine fragments. In the case of \textsuperscript{132}Sn, M1 was applied : the three transitions feeding the ground state lie above 4 MeV and could be fitted directly in the prompt $\gamma$-ray spectrum. 
\begin{figure}[h]
    \centering
    \includegraphics[width=\linewidth,clip]{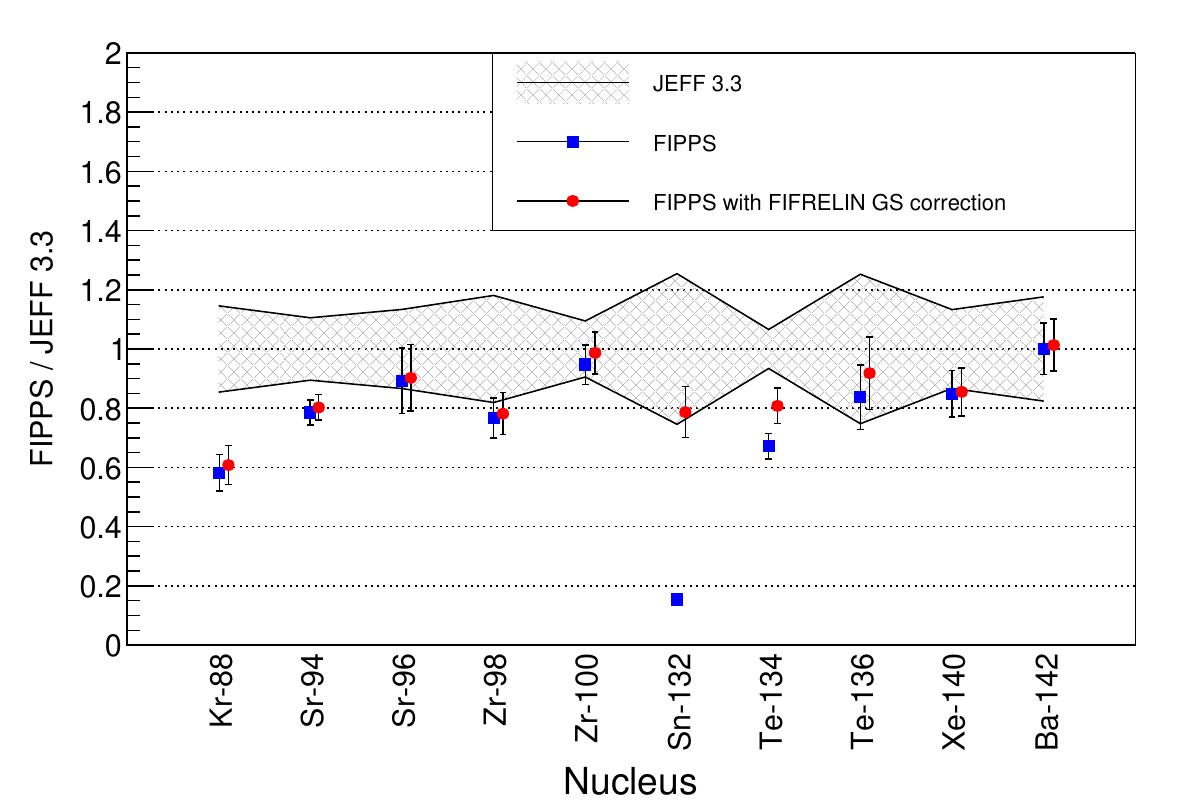}
    \caption{Absolute independent fission yields obtained in this work versus JEFF-3.3 evaluated values for a set of well produced fragments. Blue / red marks correspond to results uncorrected / corrected with FIFRELIN direct ground state production.}
    \label{fig-prompt-results}
\end{figure}

The results are shown in Fig.~\ref{fig-prompt-results}, where measured absolute fission yields are compared with JEFF-3.3 evaluated values \cite{JEFF33}. The probability of a direct ground state production, either at fission or after neutron emission, was estimated using the FIFRELIN fission fragments de-excitation code \cite{litaizeFissionModellingFIFRELIN2015}.
The overall agreement is very good, thanks to the application of many corrections to account for dead time, true summing effect, possible contamination, isomeric transitions as well as some effort on improving the precision in the absolute calibration of the spectrometer. Two exceptions remain: the lower yield of \textsuperscript{88}Kr and \textsuperscript{94}Sr, for which we do not yet have a clear explanation. The case of the double magic \textsuperscript{132}Sn is particularly noteworthy. We measure a yield of 0.107(5), a factor 5\textendash7 lower than evaluated yields: 0.74(19) in JEFF-3.3 and 0.59(2) in ENDF/B-VIII \cite{ENDFBVIII}. This indicates that \textsuperscript{132}Sn is mainly produced in its ground state, without the emission of $\gamma$ rays, in fission. This finding confirms the prediction from FIFRELIN simulations, and which find, e.g. in the case of the simulation with the CGCM level density model, a 66\% direct production of \textsuperscript{132}Sn in its ground state at scission and a 14\% direct production after neutron evaporation. For the other nuclei, such hidden ground state feeding is much weaker. 
For \textsuperscript{134}Te, a nucleus with two fewer protons than \textsuperscript{132}Sn, the same FIFRELIN simulations predict a rather large 17\% hidden production. Applying such correction leads to a result visibly closer to the JEFF-3.3 fission yield, but its significance is lower given the uncertainties.  

\section{First results on delayed gamma rays}
\label{delayed}
The high efficiency of the active-target setup for fission events (larger than 99.99 \%), the absolute calibration of the FIPPS spectrometer, and the access to delayed $\gamma$ rays make it possible, in principle, to test the combined accuracy of cumulative fission yields and decay data in a straightforward way. Indeed, the number of detected delayed $\gamma$ rays for a given transition in a nucleus depends on the accumulated fission yields of its parent nuclei and the $\gamma$-ray intensities of that transition per $\beta$ decay. In practice, however, the delayed $\gamma$ spectrum is too complex, too crowded, for reliable analysis. We prefer to use the less dense delayed $\gamma$-$\gamma$ coincidence matrix. We selected and analyzed the cleanest and/or most intense cascades of two transitions fed by $\beta$ decay in a 
small set of seven well-produced fission products listed in table \ref{tab:placeholder}.

\begin{table}[h]
    \centering
    \caption{Delayed cascades of successive transitions measured in this work}
    \label{tab:placeholder}
    \begin{tabular}{|c|c|c|c|}
    \hline
    Parents & Daughter & E\textsubscript{1} (keV) & E\textsubscript{2} (keV) \\ 
    \hline \rule{0pt}{3ex} 
    \textsuperscript{89}Rb   & \textsuperscript{89}Sr & 1248.2 & 1032.0  \\
    \textsuperscript{90}Kr   & \textsuperscript{90}Rb & 1118.7 & 539.5, 554.4  \\
    \textsuperscript{132,132m}Sb   &  \textsuperscript{132}Te & 697.1  & 974.2  \\
    \textsuperscript{132}Sn   & \textsuperscript{132}Sb  & 899.1 & 340.5  \\
    \textsuperscript{134,134m}I   & \textsuperscript{134}Xe  & 884.1 & 847.0  \\
    \textsuperscript{138,138m}Cs  & \textsuperscript{138}Ba & 462.8 & 1435.8  \\
    \textsuperscript{144}La   & \textsuperscript{144}Ce & 541.2 & 397.4  \\
    \hline
    \end{tabular}
    
\end{table}

\begin{figure}[h]
    \centering
    \includegraphics[width=\linewidth,clip]{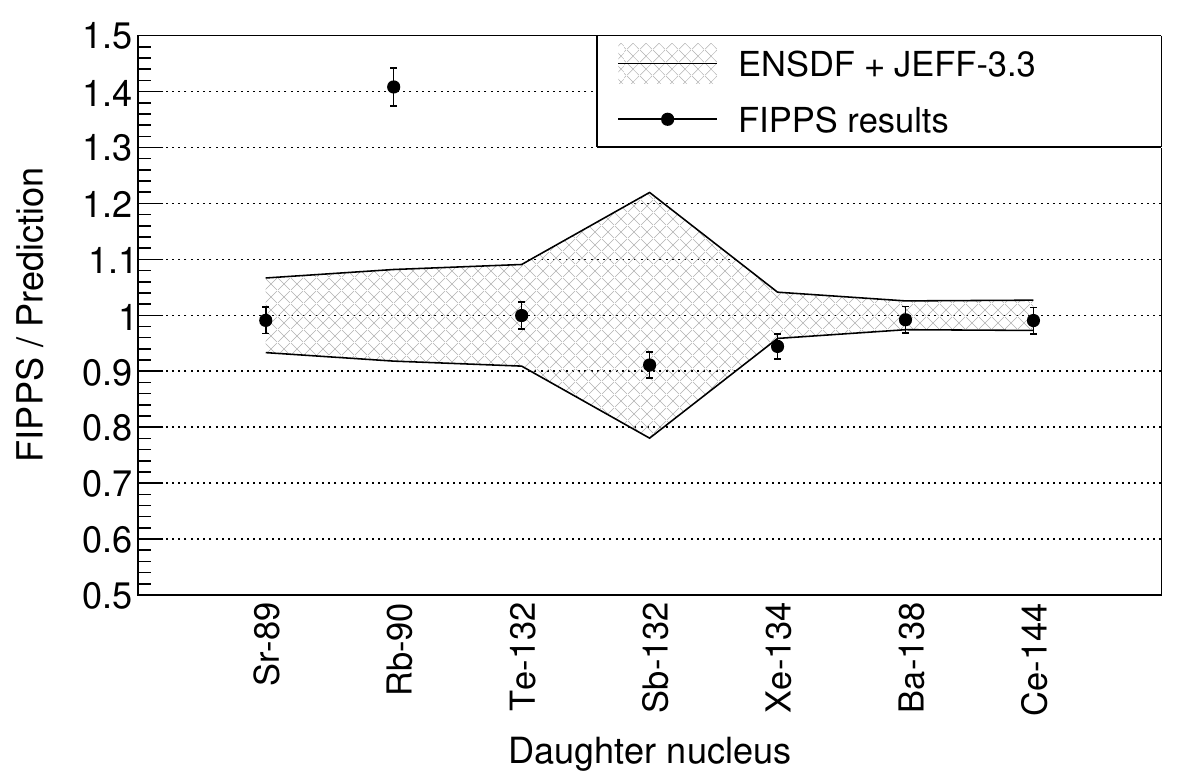}
    \caption{Comparison between the measured and the predicted values for the number of a selected delayed cascade of two transition in a nucleus per fission. See the text for more details. }
    \label{fig-delayed-results}
\end{figure}

We compared our experimental values (detected number of cascades per fission) with the expected ones from the nuclear databases on Fig.~\ref{fig-delayed-results}. Cumulative fission yields were taken from JEFF-3.3. Corrections were applied for fission products with a long lifetime relative to the measurement time using the FISPACT-II evolution code \cite{subletFISPACTIIAdvancedSimulation2017}. Decay and nuclear structure data were taken from ENSDF \cite{EvaluatedNuclearStructure}. 

Except for the decay of \textsuperscript{90}Kr to \textsuperscript{90}Rb, our results agree very well with nuclear database values and our experimental uncertainties are lower than the combined database ones. No anomaly is observed in the cascade fed by the decay of \textsuperscript{132}Sn to \textsuperscript{132}Sb. Since the cumulative fission and independent yields of \textsuperscript{132}Sn are expected to be very close according to JEFF-3.3, there is no indication that the large deficit described in section \ref{prompt} may be linked to an inaccurate fission yield in the databases. In contrast, the two cascades we measured in the decay of \textsuperscript{90}Kr are 41(3)\% and 44(3)\% higher than predicted. This discrepancy may arise from an underestimated cumulative yield of \textsuperscript{90}Kr and/or from an underestimated intensity for the 1118.7-keV transition in \textsuperscript{90}Rb in ENSDF. 
The second option is more probable as it is roughly consistent with a MTAS measurement that found a lower ground-state feeding than in the evaluated ENSDF data base: 7(1)\% instead of 29(4)\%  \cite{fijalkowskaImpactModularTotal2017}.  

\section{Conclusions}
\label{ml}
We have shown that precise absolute independent fission yields can be extracted from the prompt $\gamma$-ray data measured with the FIPPS spectrometer using an active target of \textsuperscript{235}U. This approach can become a valuable alternative when no direct technique is applicable. It is however a demanding approach that involved fitting peaks in the prompt $\gamma$-$\gamma$ coincidence matrix and $\gamma$-$\gamma$-$\gamma$ coincidence cube. It required applying several corrections, the largest one being for true summing effect and the most difficult one being for possible peak contamination, both estimated with FIFRELIN simulations. A notable outcome is the observation of a significant deficit in the yield of \textsuperscript{132}Sn, already reported in the fast-neutron-induced fission of \textsuperscript{238}U. This apparent anomaly is consistent with FIFRELIN predictions. It results from the large production of \textsuperscript{132}Sn directly in its ground state either at scission or after neutron evaporation, and thus without the emission of $\gamma$ rays. Our observation validates to some extend the excitation-energy sharing model implemented in FIFRELIN.

Delayed $\gamma$ rays measured with the same setup also proved to be valuable for identifying inconsistencies in the nuclear decay data or in the cumulative fission yields. The main result is the confirmation of an overestimation of the ground state feeding of \textsuperscript{90}Rb in evaluated decay data, which was already pointed out with MTAS measurements \cite{fijalkowskaImpactModularTotal2017}.

\bibliography{ND2025_v2} 
%
%
\end{document}